%% file: main.tex
\documentclass[journal,twosided]{IEEEtran}
\usepackage{cite}
\usepackage{graphicx}
\usepackage{color}
\usepackage[table, x11names]{xcolor}
\usepackage{booktabs}
\usepackage{multirow}
\usepackage[cmex10]{amsmath}
\usepackage{amssymb,amsfonts}
\usepackage{siunitx}
\usepackage{algorithmic}
\usepackage{textcomp}
\usepackage{lscape}
\usepackage[tight,footnotesize]{subfigure}
\usepackage{dblfloatfix} % To enable figures at the bottom of page
\usepackage{float}
\usepackage{caption}
\usepackage{breqn}
\usepackage[numbers,sort&compress]{natbib} % to make references appear in cited sequence
\begin{document}
\title{Thermal Crosstalk Analysis in RRAM Passive Crossbar Arrays}
\author{Shubham Pande,
%        Suresh Balanethiram, 
       Bhaswar Chakrabarti,
	  Anjan Chakravorty
	%}% <-this % stops a space
\thanks{S. Pande$^{1}$, B. Chakrabarti and A. Chakravorty are with the Department of Electrical Engineering, IIT Madras, Chennai 600036, India. email$^{1}$: e18d704@smail.iitm.ac.in}% <-this % stops a space 
% \thanks{S. Balanethiram is with the Department of Electronics and Communication Engineering, National Institute of Technology Puducherry, Karaikal 609609, India. email: sureshbalanethiram@gmail.com.}%
}
\maketitle
\input{abstract}
\input{intro}

\input{sim_detail}
\input{Rth_and_TC_extraction}
\input{application}
\input{conclusion}

\bibliographystyle{unsrtnat}

\bibliography{ref}
\end{document}

%% file: abstract.tex
\begin{abstract}
As the packing density of resistive random access memory (RRAM) devices increases, the effect of thermal cross-talk across the devices in a crossbar array arrangement influences their overall operation significantly. The electro-thermal effects in a densely packed RRAM crossbar can accelerate the retention and endurance degradation; hence poses a serious reliability threat. This paper systematically investigates the electro-thermal effects in passive RRAM crossbar arrays using COMSOL multi-physics simulations. Furthermore, we propose a methodology to model the thermal cross-talk effect and incorporate it in a SPICE compatible physics-based RRAM compact model. Finally, we demonstrate the impact of thermal coupling on RRAM crossbar array operation in terms of vector-matrix multiplication using calibrated SPICE simulations.     
\end{abstract}

%% file: intro.tex
\section{Introduction}
\IEEEPARstart{R}{eal} time processing and storage of vast amount of information in applications such as Big-Data and the Internet-of-Things demand high speed and high-density memory technologies with computational capacity. In this respect, resistive random access memory (RRAM) is a serious contender along with the other non-volatile memories such as phase change memories, ferroelectric memories and spin-transfer torque RAMs \cite{NVM_intro, PCM_intro,spinotronics_intro}. RRAM offers multi-bit storage per cell, sub-nanosecond switching speed, good endurance,  retention and viability of 3D integration in the back-end-of-line in standard CMOS processes. These intriguing properties bring the possibility of RRAM replacing the NAND flash in the near future. Note that RRAM-based crossbar arrays can perform vector-matrix multiplication (VMM) in a single time step, thereby, paving the way for in-memory computation.
\\
The promise for energy-efficient VMM operation using RRAM crossbar arrays has led to intense investigation of different crossbar architectures. The one transistor and one resistor (1T-1R), or the more general one selector and one resistor (1S-1R) architecture employs a selector device in series with the memory cell \cite{TED_2016_1T-1R,Nature_1T1R,Nature_1T1S}. The selector device provides a precise control over the filament growth during programming by controlling the current compliance to the RRAM cell \cite{TED_ielmini_cb_prog}. It also helps to suppress the cell leakage current, thereby offering better immunity against the half select cell disturbance \cite{selector_study}. However, the active 1T-1R crossbar arrays inevitably add a significant area overhead due to the additional selector devices and their associated bulky and (often) power-hungry peripheral circuitry \cite{selector_study}. On the other hand, the passive or one resistor (1R) RRAM crossbar arrays do not require a selector device; hence, such arrays offer significant reduction in the crossbar footprint and fabrication cost. Recent reports demonstrate the realization of up to 4k passive crossbar arrays exhibiting high
uniformity, significant yield (exceeding 99\%), and high precision of conductance tuning \cite{TED_strukov_passive,nature_strukov_passive,nature_passive_cb}. Given the inherent benefits of better scaling and lower fabrication costs, the passive RRAM crossbar arrays offer promising solutions for the next-generation in-memory computing applications. Nevertheless, heavy computational loads in such applications can lead to a high
power density and consequently, overheating and higher thermal budget.\\
This issue is expected to be more severe in passive 1R crossbar arrays due to higher levels of sneak currents. On the other hand, local temperature of a memory cell critically affects the performance of a RRAM device. This can be understood from the fact that the formation and dissolution of conductive filaments (CFs) are driven by generation and  drift/ diffusion of vacancies. Therefore, Joule heating plays a critical role in the switching dynamics of filamentary RRAMs; hence, temperature change during RRAM operation is expected to significantly influence the resistive switching process as elaborated in \cite{adm_joule_heat,TED_joule_heat,acs_nano_joule_heat,Jap_2014_tc}. When the temperature increases, the ability to hold the programmed values becomes weaker, and the RRAM conductance starts to drift away. 
Such variations in the programmed values will adversely affect the RRAM-based system's performance metrics, such as classification accuracy \cite{IEEE_micro_tc_inference}. Several prior works have tried to alleviate the impact of thermal
issues from both the algorithm and hardware perspectives \cite{DAC_2018_tc_inference,DAC_2019_tc_inference,TED_2019_tc_inference_2,TED_2019_tc_inference,HPCA_2018_tc_inference,ArXiv_2020_tc_inference,ICCAD_2020_tc_inference,IEEE_micro_tc_inference}.\\
Scaling down the feature size (F) in NiO RRAM
from a 100 nm to a 30 nm node can lead to an increase in temperature from approximately 400 to 1800 K \cite{Jap_2014_tc}. Commercial implementation of RRAM requires a high density of devices, thus making it essential to study the effect of thermal crosstalk. To this end, several works based on numerical simulation \cite{srep_3d_crossbar,oa_3d_crossbar,ted_3d_crossbar,TED_2021_3d_crossbar} and experimental studies \cite{apl_2d_crossbar,mdpi_2d_crossbar,adm_joule_heat,acs_nano_joule_heat,TED_joule_heat} have been reported in the literature. In \cite{srep_3d_crossbar,ted_3d_crossbar}, using numerical simulations, the authors demonstrated the dominant role of transient thermal effects on the reset mechanism, the impact of thermal crosstalk on RRAM retention characteristics, and the scaling potential of 3-D RRAM arrays. On similar lines, authors in \cite{oa_3d_crossbar} have proposed two types of structures, coined as the ‘thermal house' (TH), that facilitate thermal management, and studied electro-thermal effects in 3-D RRAM crossbar arrays through detailed numerical simulations. In \cite{TED_2021_3d_crossbar}, a 3-D electro-thermal numerical model is used to perform the signal and thermal integrity analysis of 3-D stacked Resistive-switching random access memory (RRAM) arrays. 
The experimental work in \cite{apl_2d_crossbar} has measured the temperature profile of TaO$_{x}$ RRAM during the operation using an ultra-fast imaging system. The maximum temperature rise at the top surface was observed to be $\approx$ 147 K when a low resistance device was biased at 1.8 V (very close to the reset voltage); however, it is expected that the internal temperature rise can be much higher than the measured temperature. Furthermore, the authors demonstrated thermal crosstalk within an array of RRAM devices via heat dissipation in a single device. It has been observed that nearby devices were affected more, thus, exemplifying the presence of thermal crosstalk in 2D crossbar arrays. The investigation carried out in \cite{mdpi_2d_crossbar} shows that switching a cell repeatedly leads to an increased Joule heating in the device. This heat is transported preferentially along the electrode metal lines,
affecting the neighboring cells positioned along the same electrode lines and causing a deterioration of
their electrical properties.\\
While it is true that thermal crosstalk will be more pronounced in 3D crossbar arrays, indeed, it is not insignificant in 2D crossbar arrays \cite{apl_2d_crossbar,mdpi_2d_crossbar,IEEE_micro_tc_inference}.
However, we have not found any study in the literature that systemically investigates the thermal crosstalk in 2-D crossbar arrays and provides a framework to incorporate it in SPICE simulations. This work attempts to investigate electro-thermal effects in passive 2-D crossbar arrays. The main contributions of this paper are summarized below.
\begin{enumerate}
    \item Through numerical simulations, we systematically investigate the thermal crosstalk in passive RRAM crossbar arrays assuming realistic boundary conditions.
    \item We detail the procedure to quantify thermal crosstalk and incorporate it in a SPICE compact model.
    \item Using a physics-based RRAM compact model, we evaluate the impact of thermal crosstalk on VMM accuracy.
\end{enumerate}

\noindent Our work will aid incorporation of thermal crosstalk in reliable SPICE simulations. This, in turn, is expected to help device and circuit engineers to design thermally aware RRAM based systems.\\
The rest of the paper is organized as follows. Section II elaborates the simulation setup, followed by a discussion on thermal resistance, thermal coupling coefficient extraction, and SPICE implementation of the thermal coupling framework in section III. In Section IV, we illustrate the impact of thermal crosstalk on RRAM crossbar array operation using VMM as an example. Finally, we conclude in Section V.

%% file: sim_detail.tex
\section{Simulation Setup Details}
\begin{figure}[t!]
\centering
\includegraphics[width=0.48\textwidth]{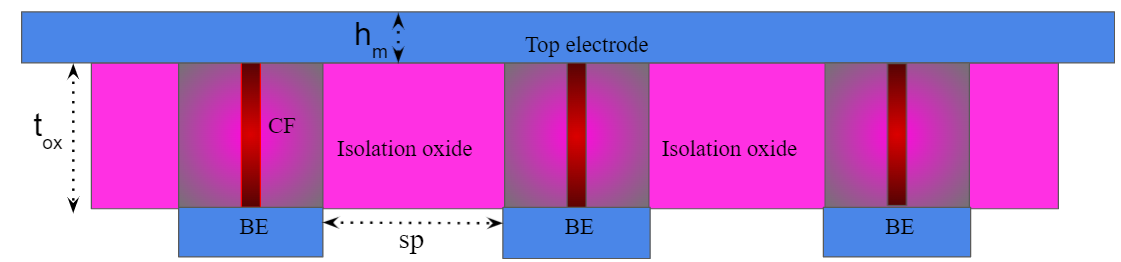}
\caption{Schematic of the structure investigated in this study.}
\label{sc_2d}
\end{figure}

Fig.~\ref{sc_2d} represents a schematic of a typical 1$\times$3 RRAM passive crossbar array where the RRAM cells are formed at the intersections of the top electrode lines and the bottom electrode lines. The switching dielectric layer also provides electrical isolation between the neighboring cells. Spacing between two adjacent cells is denoted by 'sp'. Thickness of the switching oxide layer and the metal electrodes are denoted by t$_{ox}$ and
h$_{m}$, respectively. \\ 
%%*************************************************************************
%------------------------------------------
%-------------------------------------------------------
To evaluate the thermal behavior in a crossbar array, we solve the time-dependent heat flow equation that reads 
\begin{align}
\begin{split}
        c \rho \frac{\partial T}{\partial t} &=\nabla \kappa_{th} \nabla T + \sigma |\nabla E|^2. \\
 \end{split}
 \label{heat}
 \end{align}
Here, c, $\rho$, $\kappa_{th}$, and $\sigma$ denote specific heat capacity, density, thermal conductivity, and electrical resistivity, respectively, of the material. Temperature, voltage, and the electric field are indicated by T, V, and E, respectively. Electrical field is obtained by solving the current continuity equation and then taking the gradient of V as   
\begin{align}
\begin{split}
%       &\nabla. \sigma. \nabla V = 0 \textcolor{red}{(???)}\\ 
       &E = -\nabla V.
       \label{poisson}
\end{split}
\end{align}
Thus, electro-thermal effects inside the RRAM are characterized by solving ~\eqref{heat} and ~\eqref{poisson} self-consistently using COMSOL multi-physics solver \cite{comsol}.
Electro-thermal effects mainly drive the reset process in RRAM, and maximum temperature is observed during the reset process \cite{apl_2d_crossbar,ted_3d_crossbar,srep_3d_crossbar}. Therefore, we simulate only the reset process to reduce the computational cost. In other words, we assume that a cylindrical filament already exists (formed during the prior set process) between the top and bottom electrodes as shown in Fig.~\ref{sc_2d}.   
All geometrical and material-specific parameters of the crossbar considered in this study are listed in the Table~\ref{tab:parameters}.
Please note that we ignore the possible temperature dependent variations for the material-specific parameters, such as thermal conductivity ($\kappa$), in this study. Nevertheless, if required, we can incorporate its temperature dependence through Kirchhoff's transformation technique\cite{Kirchoff_T}. 
\input{table}\\
%%*************************************************************************
%------------------------------------------
\begin{figure}[t!]
\centering
\includegraphics[width=0.5\textwidth]{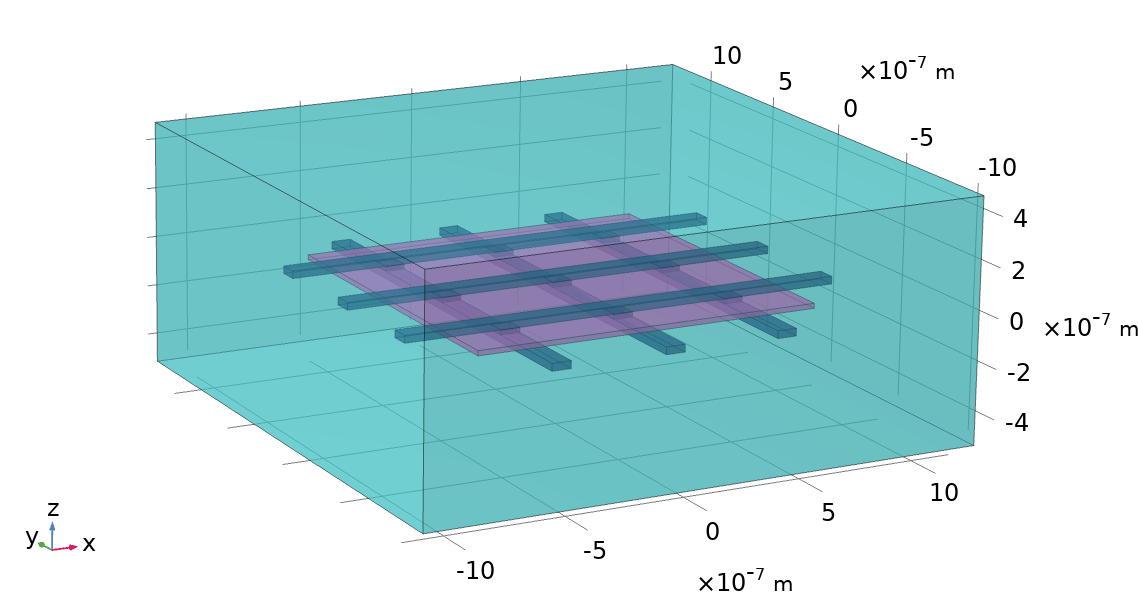}
\caption{The simulated unit cell of a 3$\times$3 passive RRAM crossbar array.}
\label{cb_2d}
\end{figure}%
%------------------------------------------
% %%*************************************************************************
\noindent 
In many thermal modeling studies the electrodes of an RRAM cell are often treated as perfect heat sinks \cite{ted_3d_crossbar,srep_3d_crossbar,oa_3d_crossbar,JAP_material_parameters}. However, experimental studies reported in \cite{apl_2d_crossbar, mdpi_2d_crossbar} suggest that local temperature of the metal electrode lines may considerably exceed the ambient room temperature. In a 2D crossbar array, electrode lines are expected to carry a significant part of heat dissipation along the direction of metallic lines due to their superior thermal conductivity. Therefore, treating electrodes as perfect heat sinks would practically ignore any thermal crosstalk as most of the heat flux would terminate on the electrodes themselves.  

% Earlier reported works\cite{ted_3d_crossbar,srep_3d_crossbar,oa_3d_crossbar} assumed the top (bottom) surface of the top (bottom) electrode was fixed at 300K (heat sink). Therefore, in the case of a 2D crossbar array, we began our analysis assuming the top (bottom) surface of the top (bottom) electrode is fixed at 300K (heat sink), and all other surfaces were considered to exhibit adiabatic boundary conditions. The RRAM cell is excited by applying the voltage signal on the top electrode, and the bottom electrode is maintained at the ground. Simulation results obtained using above said boundary conditions did not show the thermal crosstalk though the temperature of the individual cell has increased significantly (due to self-heating). This is because almost all the heat flux lines originating from the heat source (CF) could terminate on the heat sink kept on metal lines rather than flowing towards the neighboring device. For the given active layer thickness, the heat sink is closer than the adjacent RRAM devices. These simulation results are in contrast with the experimental observation reported \cite{apl_2d_crossbar,mdpi_2d_crossbar}.
 
To realize the realistic boundary conditions, we enclose the crossbar array in a very low thermal (0.05$\frac{W}{mK}$) conductivity region, denoted as the thermal house (TH). The crossbar is placed at the center of the TH, and outer surfaces of TH are kept at 300 K. The simulated unit cell of a 3 $\times$ 3 passive RRAM crossbar array is shown in Fig.~\ref{cb_2d}.

% The RRAM cell is formed at the intersection of the top and bottom metal lines, and each cell is separated from the other by the isolation oxide. We are considering only the reset process, and we assume that the cylindrical CF of equal radius exists in programmed RRAM cells. Although practically programmed RRAM cells could have different CF radius as far as this analysis is considered, we study the effect of thermal coupling for a given CF radius because one can easily extrapolate these results to the case where the CF's radius is not equal. 
% Fig.~\ref{tc_2d} shows the temperature profile along the 2$^{nd}$ row of the 3 $\times$ 3 unit cell obtained by exciting all the RRAM cells together. Here the temperature of the center RRAM cell is higher than the corner, which is attributed to thermal coupling.   
% % %------------------------------------------
% \begin{figure}[h!]
% \centering
% \includegraphics[width=0.45\textwidth]{Fig/tc.png}
% \caption{Temperature profile along the 2$^{nd}$(center) row of a simulated 3 $\times$3 crossbar array.}
% \label{tc_2d}
% \end{figure}%
% % %------------------------------------------

%% file: table.tex
\begin{table}[h!]
\centering
\begin{tabular}{|c|c|c|c|}
\hline
\rowcolor[HTML]{CBCEFB} 
                                                                      Parameters         & CF                                                        & Electrode & Oxide     \\ \hline
Width {[}m{]}                                                                  & \begin{tabular}[c]{@{}c@{}} 5e-9\\ (radius)\end{tabular} & 80e-9     &  80e-9 \\ \hline
Height {[}m{]}                                                                 & 20e-9                                                     & 30e-9     & 20e-9     \\ \hline
\begin{tabular}[c]{@{}c@{}}Electrical \\ conductivity \\ {[}$\frac{S}{m}${]}\end{tabular} & 7e3 & 1.23e5 & 7e-7 \\ \hline
\begin{tabular}[c]{@{}c@{}}Thermal \\ conductivity \\ {[}$\frac{W}{mK}${]}\end{tabular}   & 22  & 22     & 0.5  \\ \hline
\begin{tabular}[c]{@{}c@{}}Heat capacity \\ {[}$\frac{J}{kgK}${]}\end{tabular} & 445                                                       & 445       & 286       \\ \hline
\begin{tabular}[c]{@{}c@{}}Density \\ {[}$\frac{Kg}{m^3}${]}\end{tabular}      & 8.9e3                                                     & 8.9e3     & 9.68e3    \\ \hline
\end{tabular}
\caption{Parameters used for this study. The electrical and thermal parameters mentioned in \cite{JJAP_material_parameters,srep_3d_crossbar,ted_3d_crossbar,JAP_material_parameters} are used as a reference for this work.}
\label{tab:parameters}

\end{table}

%% file: Rth_and_TC_extraction.tex
\section{Thermal Parameters Extraction and Model Framework}
\subsection{Thermal resistance}
\begin{figure}[t!]
\centering
\includegraphics[width=0.27\textwidth]{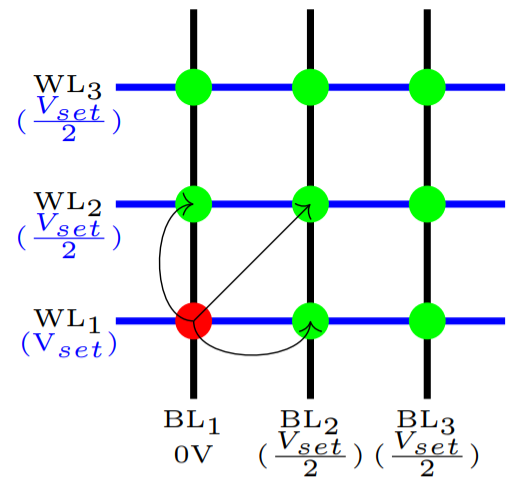}
\caption{Schematic of a 3$\times$3 crossbar array. The red circle shows the excited RRAM cell in LRS}.
\label{cb_sch}
\end{figure}

\begin{figure}[h!]
\centering
\includegraphics[width=0.5\textwidth]{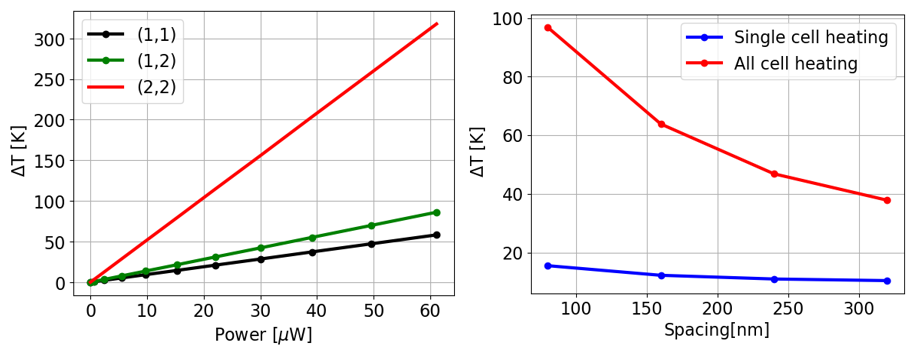}
\caption{(Left) P$_{diss}$-dependent $\Delta T$ for a fixed spacing (sp). Here, cell (2,2) is acting as the heat source. (Right) Spacing dependent $\Delta T$ at fixed P$_{diss}$.}
\label{pdiss_vs_tc}
\end{figure}

\begin{figure}[h!]
\centering
\includegraphics[width=0.5\textwidth]{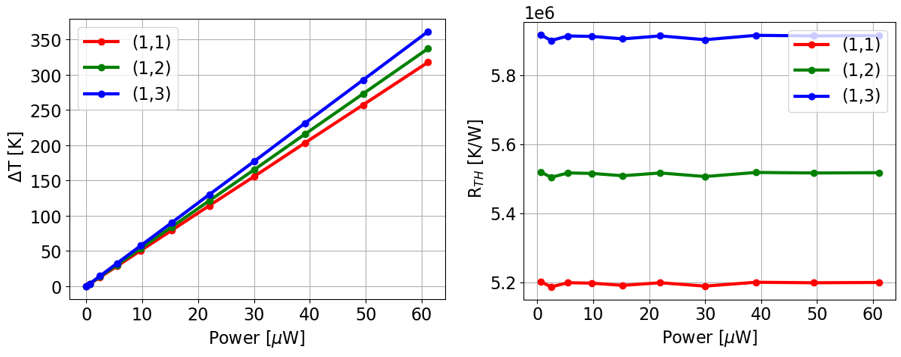}
\caption{P$_{diss}$-dependent (Left) $\Delta T$ and (Right) R$_{th}$ for individually excited RRAM cells.}
\label{pdiss_vs_T}
\end{figure}

Thermal resistance (R$_{th}$) of an individual RRAM cell in a crossbar array would depend on its relative position from the heat sink and the number of other cells to which it is connected. Therefore, we begin extracting R$_{th}$ corresponding to each cell. Fig.~\ref{cb_sch} shows the biasing scheme implemented for accessing individual RRAM cells. RRAM cells in low resistance state (LRS) are treated as a heat source and are shown in red color. While extracting the R$_{th}$ of an individual cell, we assume that the particular cell is in LRS and all other cells are in HRS. This procedure ensures that temperature rise at the excited cells is only due to self-heating. The steady-state temperature rise ($\Delta T$) due to self-heating of the excited RRAM cell can be written as
\begin{equation}
    \Delta T = T - T_{amb} = P_{diss}\times R_{TH}
\end{equation}
with
\begin{equation}
    P_{diss} = V_{applied}\times I_{(TE,BE)}
\end{equation}
where T$_{amb}$, R$_{TH}$ and P$_{diss}$ are ambient temperature, thermal resistance, and power dissipation, respectively.\\
% %----------------------------------------------------------------
The time taken by the crossbar array to reach a thermal steady-state (t$_{s}$) depends on the size of the system and its thermal properties. A larger crossbar array would have higher t$_{s}$ due to increased thermal mass \cite{srep_3d_crossbar}. Duration of the read/ write pulses in moderately large arrays is typically in the range of a few hundreds of $\mu$s \cite{TED_ielmini_cb_prog,nature_strukov_passive}. In this study, the pulse width of the applied voltage signal is 10$\mu$s, and it is sufficient to ensure that the crossbar system has reached a thermal steady state. The radius of the conductive filament is fixed at 5 nm, and the amplitude of the voltage pulse is varied to realize the power sweep. Assuming RRAM LRS resistance is in the range k$\Omega$s, and the cell is operating at a current compliance of a few $\mu$As, resulting power dissipation in LRS would be in $\mu$Ws.
We plot the temperature at the center of CF (maximum temperature) unless otherwise stated.\\
Fig.~\ref{pdiss_vs_tc} (left) shows the $\Delta T$ variation with respect to the power dissipated (P$_{diss}$) for a spacing (sp) equal to 400 nm. Here the RRAM cell located at 2$^{nd}$ row and 2$^{nd}$ column (2,2) is treated as a heat source (single heat source scenario). Due to symmetry, $\Delta T$ in (1,1) is the same as that at (1,3), (3,1), and (3,3) cells. Similarly, $\Delta T$ in (1,2), (2,1), (2,3), and (3,2) would be the same. Passive temperature rise in the neighboring cell due to the self-heating of the (2,2) cell is around 50 K for the given spacing, and for the range of P$_{diss}$. Fig.~\ref{pdiss_vs_tc} (right) shows $\Delta T$ variation of (2,2) cell with respect to spacing at P$_{diss}$ equal to 2.45 $\mu$W. When only the (2,2) cell is in LRS (heat source), the maximum $\Delta T$ is 18 K for a spacing of 80 nm. However, if we consider the worst-case scenario where all cells are in LRS (heat source), the maximum $\Delta T$ could reach around 100 K. Thus, thermal crosstalk becomes prominent as we reduce the spacing between the two cells.\\  
%------
In Fig.~\ref{pdiss_vs_T} (left), we show the dissipated power dependent $\Delta T$ obtained by exciting one cell at a time. It is observed that the 
steady-state temperature varies with the position of the RRAM cell in a crossbar array. This is because the distances between the heat source (the programmed RRAM cell) and the thermal dissipation boundaries might differ from cell to cell. For example, the generated heat at (2,2) cell can be quickly dissipated towards neighboring cells via the electrodes. However, corner cells would find it relatively difficult to dissipate heat via electrode lines and thus show higher temperature rise than the cell located at the center for the given P$_{diss}$. Fig.~\ref{pdiss_vs_T} (right) shows extracted R$_{th}$ of an individual cell. Please note that R$_{th}$ is constant with respect to power as we are ignoring temperature dependence of thermal conductivity.   

% %------------------------------------------
% %---------------------------------------------------------

% %------------------------------------------

% %------------------------------------------
\subsection{Thermal coupling coefficient}
\begin{figure}[h!]
\centering
\includegraphics[width=0.5\textwidth]{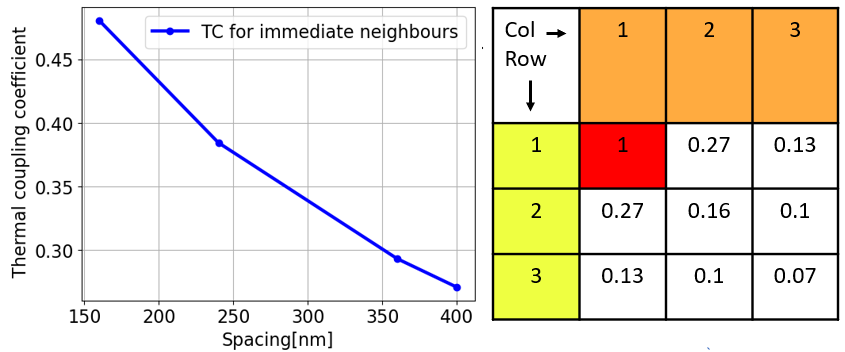}
\caption{(Left) Spacing dependent TC associated with immediate neighboring cells. (Right) TC matrix when cell (1,1) is heating alone.}  
\label{spacing_vs_tc_matrix}
\end{figure}

% %------------------------------------------
To quantify the extent of thermal crosstalk, we define a thermal coupling coefficient (TC). It is characterized as the ratio of the passive temperature rise in cell\#1 due to the self-heating of cell\#2. The detailed procedure of extracting the thermal coupling coefficient is given below. 
The steady-state temperature rises at an RRAM cell\#1 in response to its heat dissipation can be expressed using the device’s dissipated power and thermal resistance as
\begin{equation*}
    \Delta T_{11} = P_{diss1} \times R_{th1}.
\end{equation*}
The additional rise, caused by a second RRAM cell\#2, in terms of it's respective power dissipation will be
\begin{equation*}
    \Delta T_{12} = P_{diss2} \times R_{th12}
\end{equation*}
with $R_{th12}$ being a thermal resistance between RRAM cell\#1 and RRAM cell\#2. The total temperature rise at RRAM cell\#1 then can be written as
\begin{align*}
 \Delta T_{1} &= \Delta T_{11} + \Delta T_{12},
\end{align*}
\begin{align}
 \Delta T_{1} &= P_{diss1} \times R_{th1} +  P_{diss2} \times R_{th12}.
 \label{T1}  
\end{align}
The dissipated power P$_{diss2}$ of RRAM cell\#2 can be expressed in terms of its $\Delta T_{22}$ and thermal resistance R$_{th2}$ as
\begin{equation}
  P_{diss2} = \frac{\Delta T_{22}}{R_{th2}}.
  \label{pdiss2}
\end{equation}
Inserting \eqref{pdiss2} in \eqref{T1} yields
\begin{align*}
 \Delta T_{1} &= P_{diss1} \times R_{th1} + \frac{R_{th12}}{R_{th2}} \Delta T_{22}. 
\end{align*}
The thermal coupling coefficient
\begin{align*}
 c_{12} &= \frac{R_{th12}}{R_{th2}}
 \label{Tc1}  
\end{align*}
characterises the impact of the self-heating of cell\#2 on temperature rise of cell\#1. Rearranging \eqref{T1} and setting the power dissipation term P$_{diss1}$ to zero result in the static thermal coupling coefficient as
\begin{align*}
 c_{12} &= \frac{\Delta T_{1}}{\Delta T_{22}}\Big\vert_{P_{diss1}=0}.
\end{align*}
Thus, the coupling coefficient c$_{12}$ is the quotient of the temperature rises at cell\#1 and \#2 for the case where cell\#1 is switched off, and cell\#2 is switched on. One can easily generalize this approach and apply it to a crossbar-based system to express
\begin{align*}
\Delta T_{1n} &= R_{th1n}P_{dissn} + \sum_{m = 1 (m \neq n)}^{i}c_{nm} \Delta T_{mm}, \\
c_{nm} &= \frac{\Delta T_{n}}{\Delta T_{mm}}\Big\vert_{P_{dissm}=on}.
\end{align*}
Introducing a thermal coupling matrix for i$^{th}$ number of cells, one can write.
\[
\begin{bmatrix}
    \Delta T_{1} \\
    \Delta T_{2} \\
    \vdots       \\
    \Delta T_{i} 
\end{bmatrix}
=
\begin{bmatrix}
    1 & c_{12} & c_{13} & \dots  & c_{1i} \\
    c_{21} & 1 & c_{23} & \dots  & c_{2i} \\
    \vdots & \vdots & \vdots & \ddots & \vdots \\
    c_{i1} & c_{i2} & c_{i3} & \dots  & 1
\end{bmatrix}
\begin{bmatrix}
    \Delta T_{11} \\
    \Delta T_{22} \\
    \vdots       \\
    \Delta T_{ii} 
\end{bmatrix}
\]
Note that the main diagonal elements are unity (c$_{11}$, c$_{22}$...c$_{ii}$) because they describe the influence of the respective sources on themselves. The other elements in the matrix can be found by exploiting the symmetry in the crossbar structure,
\begin{align*}
    c_{ab} = c_{ba}. 
\end{align*}
% The temperature rise at excited cell can induce passive temperature rise at neighboring devices.
The thermal coupling matrix associated with cell (1,1) at a spacing equal to 400 nm is shown in Fig.~\ref{spacing_vs_tc_matrix} (right). 
The neighboring cells situated along one of the electrodes common with the heated device experience maximum thermal coupling.
The cells with no electrode lines in common with the heated cell also experience non-negligible thermal coupling, and its impact decreases as we move away from the heating source.
Fig.~\ref{spacing_vs_tc_matrix} (left) shows the TC associated with the immediate neighboring RRAM cell for varying spacing. For example, TC associated with cells (1,2) and (2,1) when RRAM cell (1,1) is the heat source. For spacing below 160 nm, the value of TC can be higher than 0.5. 
% %------------------------------------------
% %------------------------------------------
The extent to which neighboring devices can see the impact of the thermal crosstalk would depend on factors such as the spacing between two cells, the thermal conductivity of metal lines, the width, and thickness of the metal lines \cite{mdpi_2d_crossbar}. Moreover, the thermal and electrical properties of the material used in the crossbar system depend on various factors such as growth conditions, thickness, etc. Hence, accurately calibrating the simulation setup for a specific technology is quite difficult and time consuming. Nevertheless, TC values reported here have been obtained using reasonable parameter values, and hence, can be treated as indicative instead of being specific to a certain technology.

\subsection{Modeling framework}
\begin{figure}[h!]
\centering
\includegraphics[width=0.4\textwidth]{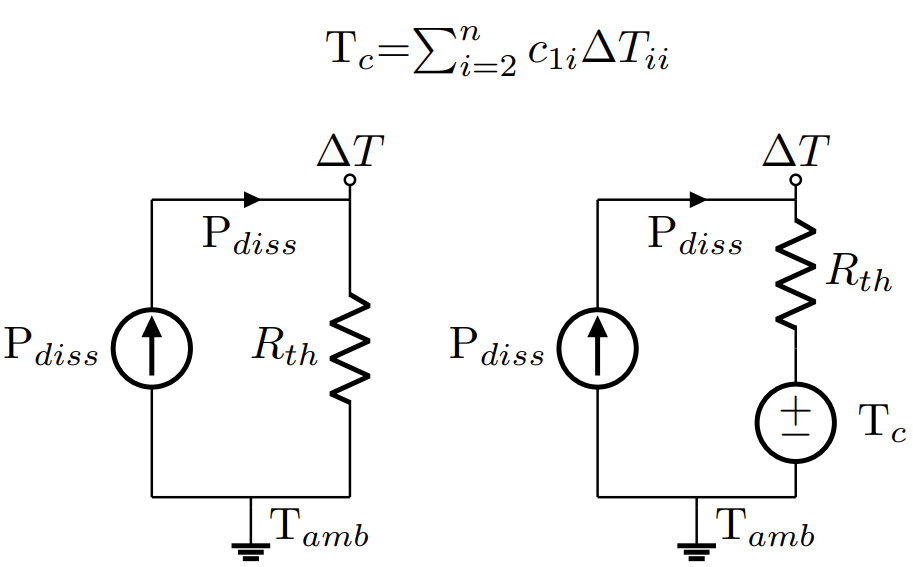}
\caption{(Left) Steady-state thermal network for self-heating. (Right) Modified self-heating thermal network incorporating thermal coupling.}   
\label{tc_nw}
\end{figure}

Conventionally temperature rise due to the self-heating is modeled by R$_{th}$ in series with a current source of value P$_{diss}$ as shown in Fig.~\ref{tc_nw} (left). The additional temperature rise due to thermal coupling can be incorporated in the self-heating thermal network by adding a voltage-dependent-voltage-source in series with thermal resistance as shown in Fig.~\ref{tc_nw}. This completes the steady state thermal sub-circuit of RRAM in a crossbar system.
% % %------------------------------------------
% % %------------------------------------------

%% file: application.tex
\section{Impact of thermal crosstalk on vector-matrix multiplication}
\begin{figure}[h!]
\centering
\includegraphics[width=0.5\textwidth]{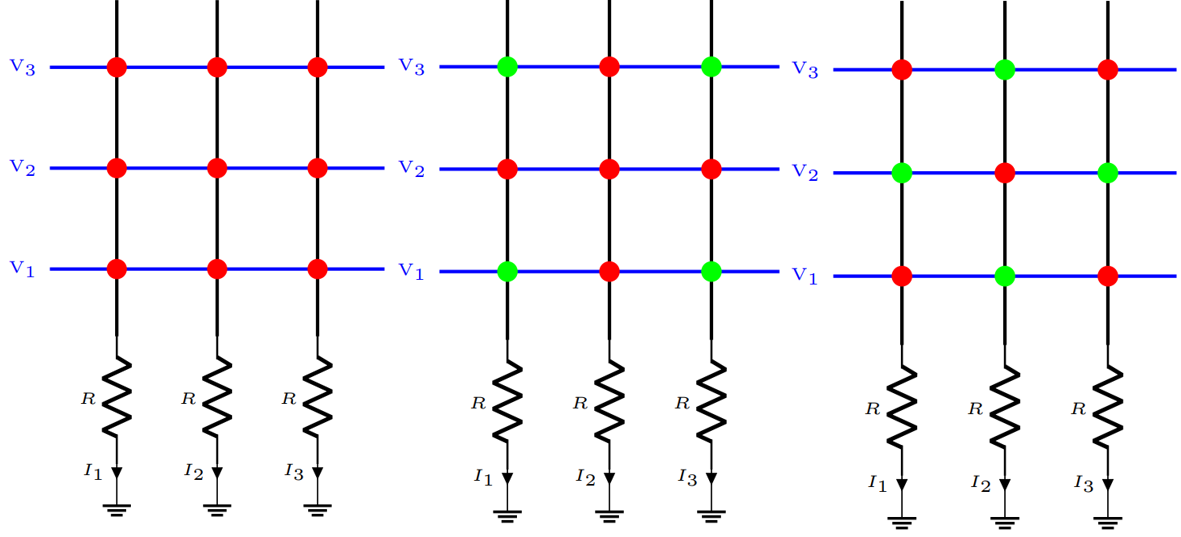}
\caption{Schematic of a 3$\times$3 passive crossbar array used in inference mode. (Left) All LRS (Center) Case A (Right) Case B. The Red (green) circle denotes the cell in LRS (HRS).}   
\label{inference}
\end{figure}

\begin{figure}[h!]
\centering
\includegraphics[width=0.5\textwidth]{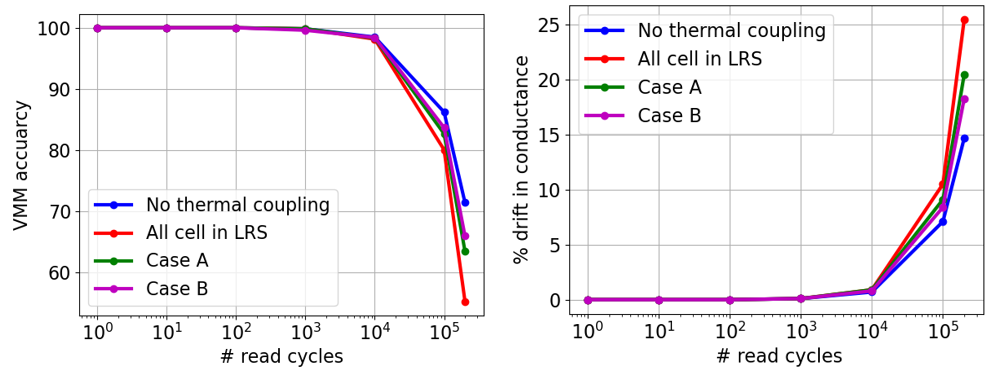}
\caption{(Left) VMM accuracy and (Right) \% drift in RRAM conductance value as a function of the number of read cycles.}   
\label{accuarcy}
\end{figure}

To illustrate the effect of thermal crosstalk on vector-matrix multiplication (VMM) operation of a crossbar, we use a 3$\times$3 passive crossbar array operating in inference mode as shown in Fig~\ref{inference}. We incorporate our thermal crosstalk framework in the physics-based SPICE compatible RRAM model reported in \cite{GIG_model}. We simulate three scenarios, including the worst case where all RRAM cells are in LRS. During inference, RRAM cells are excited by a pulse signal having an amplitude of 0.3 V and pulse width of 100 $\mu$s. We define VMM accuracy as\\
\begin{align*}
Accuracy = \Big(1- \bigg| \frac{I_{actual}-I_{ideal}}{I_{actual}+I_{ideal}}\bigg|\Big)\times100    
\end{align*}
where I$_{ideal}$ is the analytically calculated total column current and I$_{actual}$ is the current obtained from SPICE simulation using the RRAM model.
% % %------------------------------------------
% % %------------------------------------------
Each time the neural network performs the VMM, the V$_{read}$ as the input vector continues to be applied to the RRAM device. The accumulative stress induced by the V$_{read}$ may disturb the conductance of the RRAM from the read disturbance perspective, which will eventually trigger the switching operation of the RRAM. As a result, the inference accuracy of the neural network is expected to decrease \cite{ted_v_read,IRPS_v_read,ESSDERC_v_read}. Fig.~\ref{accuarcy} (left) compares the accuracy loss in VMM operation in the presence of thermal coupling. The total current in column-2 is used for reference in this case as it is more prone to thermal coupling. Due to the presence of thermal coupling, significant loss in accuracy has been observed. Furthermore, thermal crosstalk may accelerate the resistance drift and degrade crossbar-based system operation, as shown in Fig.~\ref{accuarcy} (right). Conductance drift from programmed value is calculated using
\begin{align*}
\% Drift = \frac{G_{prog} - G}{G_{prog}}\times100
\end{align*}
where G$_{prog}$ is the programmed conductance value.  
% % %------------------------------------------
% % %------------------------------------------
If LRS cells happen to be positioned along the same metal lines, it would affect the VMM accuracy as heat can dissipate into neighboring cells through metal lines; case A and case B are used to illustrate this scenario. Effect of thermal cross talk on VMM accuracy and conductance drift is negligible until 10$^{4}$ inference cycles. However, accuracy starts to degrade significantly after the 10$^{4}$ cycle. This can be understood by considering the positive feedback between voltage stress and temperature. Due to the continuously applied voltage stress, conductance starts to increase, which results in an increase in temperature due to Joule heating which further accelerates the conductance drift. With respect to no thermal coupling scenario, the additional VMM accuracy degradation is around 8$\%$, 11$\%$, 23$\%$ after 2$\times$10$^{5}$ inference cycles in case of B, Case A, and all cells in the LRS scenario, respectively. 

%% file: conclusion.tex
\section{Conclusion}
In summary, we have performed numerical simulations of passive 2-D RRAM crossbar arrays considering realistic boundary conditions. A step-by-step procedure is presented to calculate the thermal resistance and thermal coupling coefficients associated with individual RRAM cells. We have incorporated the thermal coupling framework in a compact model. Finally, we have evaluated the impact of thermal coupling in terms of accuracy degradation of vector matrix multiplication operation. 

\section*{ACKNOWLEDGMENT}
This work is partially funded by Science and Engineering Research Board (SERB), a statutory body of the Department of Science and Technology, Government of India through the project SRG/2020/001126 and by the Advanced Memory and Computing Group at the Indian Institute of Technology Madras under Institute of Eminence (IoE) scheme.